\documentclass[conference]{IEEEtran}
\IEEEoverridecommandlockouts
\usepackage{cite}
\usepackage{amsmath,amssymb,amsfonts}
\usepackage{amsthm}
\usepackage{graphicx}
\usepackage{textcomp}
\usepackage{xcolor}
\usepackage{algorithm}
\usepackage{algpseudocode}
\usepackage{graphicx}
\usepackage{float}
\usepackage{booktabs}
\usepackage{colortbl}
\usepackage{tabularray}
\usepackage{lipsum,adjustbox}
\usepackage{url}
\usepackage{cleveref}
\usepackage{xurl}
\crefname{figure}{Fig.}{Figs.} 
\def\BibTeX{{\rm B\kern-.05em{\sc i\kern-.025em b}\kern-.08em
    T\kern-.1667em\lower.7ex\hbox{E}\kern-.125emX}}

\newcommand{\PRIMEproject}{PRIME project} 
\newcommand{\PRIMEprojects}{PRIME project's } 
\newcommand{\PRIME}{PRIME } 

\makeatletter
\newcommand{\linebreakand}{%
  \end{@IEEEauthorhalign}%
  \hfill\mbox{}\par
  \mbox{}\hfill\begin{@IEEEauthorhalign}
}
\makeatother

\begin{document}

\title{Quantifying the Cross-sectoral Intersecting Discrepancies within Multiple Groups Using Latent Class Analysis Towards Fairness
}


\author{\IEEEauthorblockN{Yingfang Yuan\textsuperscript{*}}
\IEEEauthorblockA{\textit{School of Mathematical} \\
\textit{and Computer Sciences} \\
\textit{Heriot-Watt University}\\
Edinburgh, United Kingdom \\
y.yuan@hw.ac.uk}
\and
\IEEEauthorblockN{Kefan Chen\textsuperscript{*}}
\IEEEauthorblockA{\textit{School of Mathematical} \\
\textit{and Computer Sciences} \\
\textit{Heriot-Watt University}\\
Edinburgh, United Kingdom \\
kc2039@hw.ac.uk}
\and
\IEEEauthorblockN{Mehdi Rizvi}
\IEEEauthorblockA{\textit{School of Mathematical} \\
\textit{and Computer Sciences} \\
\textit{Heriot-Watt University}\\
Edinburgh, United Kingdom \\
s.rizvi@hw.ac.uk}
\linebreakand
\IEEEauthorblockN{Lynne Baillie}
\IEEEauthorblockA{\textit{School of Mathematical} \\
\textit{and Computer Sciences} \\
\textit{Heriot-Watt University}\\
Edinburgh, United Kingdom \\
l.baillie@hw.ac.uk}
\and
\IEEEauthorblockN{Wei Pang\textsuperscript{$\dagger$}}
\IEEEauthorblockA{\textit{School of Mathematical} \\
\textit{and Computer Sciences} \\
\textit{Heriot-Watt University}\\
Edinburgh, United Kingdom \\
w.pang@hw.ac.uk}
}
\maketitle

\begin{abstract}
The growing interest in fair AI development is evident. The ``Leave No One Behind'' initiative urges us to address multiple and intersecting forms of inequality in accessing services, resources, and opportunities, emphasising the significance of fairness in AI. This is particularly relevant as an increasing number of AI tools are applied to decision-making processes, such as resource allocation and service scheme development, across various sectors such as health, energy, and housing. Therefore, exploring joint inequalities in these sectors is significant and valuable for thoroughly understanding overall inequality and unfairness. This research introduces an innovative approach to quantify cross-sectoral intersecting discrepancies among user-defined groups using latent class analysis. These discrepancies can be used to approximate inequality and provide valuable insights to fairness issues. We validate our approach using both proprietary and public datasets, including both EVENS and Census 2021 (England \& Wales) datasets, to examine cross-sectoral intersecting discrepancies among different ethnic groups. We also verify the reliability of the quantified discrepancy by conducting a correlation analysis with a government public metric. Our findings reveal significant discrepancies both among minority ethnic groups and between minority ethnic groups and non-minority ethnic groups, emphasising the need for targeted interventions in policy-making processes. Furthermore, we demonstrate how the proposed approach can provide valuable insights into ensuring fairness in machine learning systems.
\end{abstract}


\section{Introduction}
\label{sec:introduction}
\renewcommand{\thefootnote}{\fnsymbol{footnote}}

\footnotetext{\textsuperscript{*}Equal contribution}
\footnotetext{\textsuperscript{$\dagger$}Corresponding author}
\renewcommand{\thefootnote}{\arabic{footnote}}

The “Leave No One Behind” (LNOB) principle emphasises the importance of addressing multiple, intersecting inequalities that harm individuals' rights \cite{unsdg2022universal}. Intersecting inequality refers to the compounded disadvantages that arise from both the overlap of marginalised social categories (e.g., being female and living with a disability) and the intersection of multiple, mutually reinforcing dimensions of exclusion (e.g., deprivation in both health and education) \cite{arciprete2022intersecting}. Meanwhile, the increasing adoption of AI tools in decision-making processes across various sectors, including health \cite{lysaght2019ai}, energy \cite{danish2023ai}, and housing \cite{chan2017evidence}, underscores the urgency of ensuring fairness in their design and implementation \cite{mehrabi2021survey}, as unfair AI systems may risk deepening existing inequalities.

AI fairness research spans various sectors, including healthcare \cite{cirillo2020sex,celi2022sources,byrne2021reducing}, finance \cite{zhang2019fairness}, and education \cite{fenu2022experts}. However, research on cross-sectoral intersecting AI fairness remains limited. The term ``cross-sectoral intersecting" refers to the interaction and overlap of multiple sectors, such as healthcare, housing, and energy.

To address this gap, we propose quantifying cross-sectoral intersecting discrepancies between groups. These discrepancies refer to differences in user profiles across groups and can serve as a proxy for underlying inequalities, providing valuable insights for stakeholders. Additionally, the quantified discrepancies in data can offer insights into AI fairness when the data is used to train models. According to the LNOB principle of equal opportunities, everyone should ideally have equal access to public services and resources without discrepancies. We prefer the term ``discrepancy'' over ``disparity'' or ``difference'' because it suggests an unexpected difference.

\begin{figure}
\centering
\includegraphics[width=6cm]{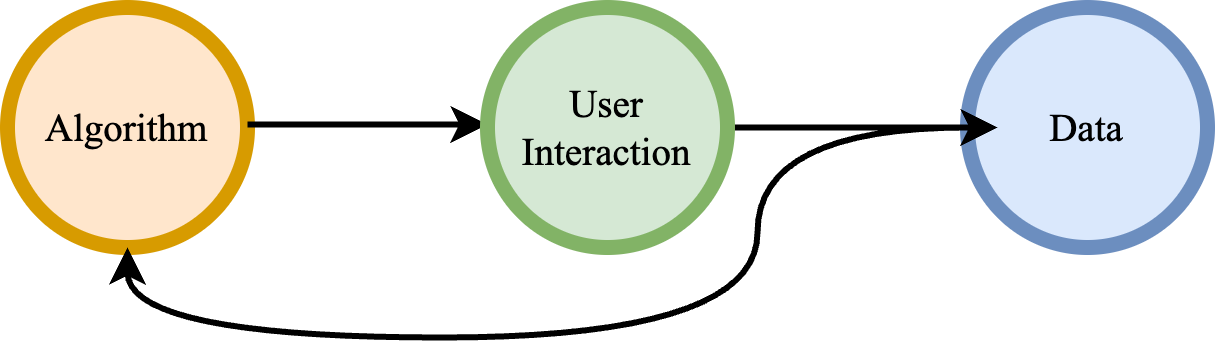}
\caption{The loop of bias placed in the data, algorithm, and user interaction feedback \cite{mehrabi2021survey}.}
\label{fig:bias_loop}
\end{figure}

\textbf{Bias and Fairness} Recent studies highlight concerns that AI-supported decision-making systems may be influenced by biases \cite{leslie2021does}, which can unfairly impact vulnerable groups, such as ethnic minorities, emphasising the need for research on AI system fairness \cite{morley2020ethics}. A primary obstacle in advancing and implementing fair AI systems is the presence of bias \cite{ferrara2023fairness}. In AI, bias can originate from various sources, including data collection, algorithmic design, and user interaction, as illustrated in \Cref{fig:bias_loop}.

\begin{figure}
\centering
\includegraphics[width=9cm]{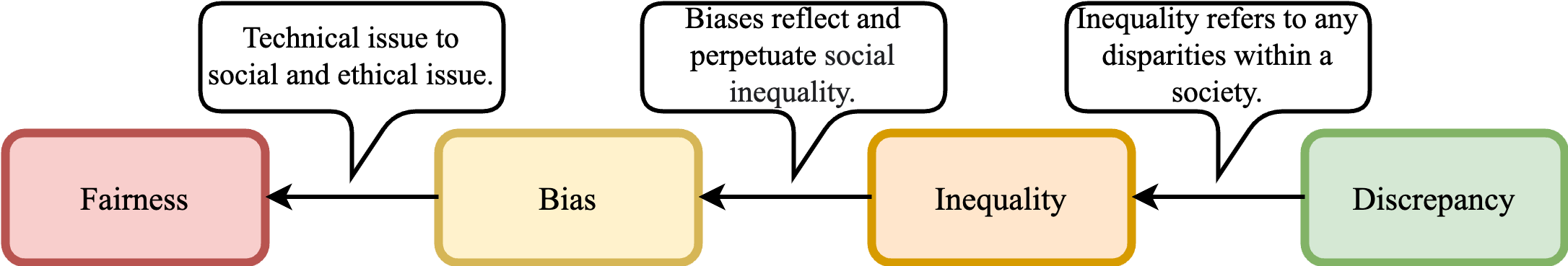}
\caption{The correlations between fairness, bias, inequality, and discrepancy in the context of AI \cite{Team_2025, ferrara2023fairness}.}
\label{fig:discrepancy}
\end{figure}

In particular, most AI systems rely on data for training and prediction. This close connection means that any inherent biases in the training data can be propagated and embedded into the AI systems, leading to biased predictions, the so called bias-in and bias-out. Even if the data itself is not inherently biased, algorithms can still exhibit biased behaviour due to inappropriate design and configuration choices. These biased outcomes can influence AI systems in real-world applications, creating a feedback loop where biased data from user interactions further trains and reinforces biased algorithms, resulting in a vicious cycle \cite{mehrabi2021survey}.

Bias stemming from data is a crucial factor affecting fairness, as inappropriate handling of it may trigger a cascade of other biases, exacerbating fairness issues. 



\textbf{Discrepancy} in data may indicate inequalities and lead to biases and unfairness, given that individuals should ideally be treated equally in ideal situations. \Cref{fig:discrepancy} illustrates the correlations between fairness, bias, inequality, and discrepancy. Essentially, fairness is indirectly linked with discrepancy, and discrepancy can contribute to unfairness. The difference between fairness and bias is that the former can be viewed as a technical issue, while the latter can be viewed as a social and ethical issue \cite{ferrara2023fairness}. Furthermore, bias is a problem caused by historical and current social inequality \cite{Team_2025}, and inequality can manifest as discrepancies. \Cref{fig:discrepancy} starts with discrepancy and moves through inequality and bias to fairness. Therefore, \textit{our research focuses on quantifying cross-sectoral intersecting  discrepancies among different groups, with an aim to uncover insights or patterns related to inequality, bias, and AI fairness}.

\textbf{Background and Motivation} Currently, there is limited research focusing on quantifying discrepancies. Most recent research on quantifying bias and/or inequality primarily revolves around resource allocation strategies and generally relies on objective data (e.g., \cite{wu2022quantifying}). However, these approaches have limitations and face challenges in effectively assessing and measuring bias or discrepancy in datasets unrelated to resource allocation. For example, in social sciences, much data is collected through questionnaires, which often include binary, categorical, or ordinal data types related to subjective responses and user experiences. These questionnaires may cover various aspects, resulting in intersecting and cross-sectoral data of high dimensions. Analysing data based on no more than two dimensions or sectors may overlook important information or patterns. Therefore, we believe that quantifying cross-sectoral intersecting  discrepancies is valuable, as it can provide comprehensive insights.

The \textit{\PRIME}project aims to establish safer online environments for minority ethnic (ME) populations in the UK. Its survey questionnaire covers five key aspects: demography, energy, housing, health, and online services. Notably, the data collected in this context does not directly pertain to resource allocation, making it challenging to explicitly define and detect bias within the data using current methods, despite the presence of discrepancies. These discrepancies may arise from various factors, including culture, user experience, and discrimination, potentially contributing to bias or unfairness. This research is mainly motivated by the \textit{\PRIME}project, so the datasets we used primarily cover the health, energy, and housing sectors, with our research targeting ME groups.

In our preliminary research (see Appendix \ref{appendix:prime}) in England, based on a \textit{\PRIME}project survey question regarding health and digital services\footnote{Question 21: Which of the following concerns do you have about communicating with your general practice (GP) through apps, websites, or other online services?}, we observed a notable discrepancy in the Chinese group: 30.16\% lacked English proficiency and 26.98\% struggled to use the online system, while most other ethnic groups reported less or no concerns. These discrepancies, stemming from cultural differences, user experiences, or discrimination, contribute to inequality and may affect AI fairness. For instance, an AI system might inappropriately assume that only Chinese individuals require English language support, thereby neglecting other ME groups who may also need assistance. 


However, investigating multiple and cross-sectoral questions simultaneously is challenging. For instance, the \textit{\PRIME} project's data contains similar questions for the energy and housing sectors, and current methods struggle to analyse these sectors jointly. Therefore, we propose an approach to quantify intersecting and cross-sectoral discrepancies for multiple ethnic groups by leveraging latent class analysis (LCA) \cite{morin2023stepmix}.

LCA is a popular method in social science \cite{collins2009latent} because it can identify latent groups within a population based on observed characteristics or behaviours. LCA offers a flexible framework for exploring social phenomena and integrating with other analytical techniques. In this research, we use LCA to cluster intersecting and cross-sectoral data, encompassing questions across the health, energy, and housing sectors. This method enables us to derive latent classes and outcomes, with each class describing a distinct cross-sectoral user profile. This approach moves beyond defining user classes solely based on individual questions. More details of our proposed approach are presented in \Cref{sec:method}, with experiments and results reported in \Cref{sec:exp}.

The main contributions of this research are as follows: (1) we propose a novel and generic approach to quantify intersecting and cross-sectoral discrepancies between user-defined groups; (2) our findings reveal that ME groups cannot be treated as a homogeneous group, as varying discrepancies exist among them; and (3) we demonstrate how the proposed approach can be used to provide insights to AI fairness.

\section{Related Work}
\textbf{Quantifying and improving AI fairness} As AI technologies are used more and more frequently in real life, people's concerns about the ethics and fairness of AI have always existed, especially when AI is increasingly used in problems with sensitive data \cite{trocin2023responsible}. Morley et al.~\cite{morley2020ethics} and Garattini et al.~\cite{garattini2019big} noticed that an algorithm ``learns'' to prioritise patients it predicts to have better outcomes for a particular disease. And they also noticed that AI models have discriminatory potential when facing ME groups on health. Therefore, people are paying more and more attention on the impact and mitigating methods of AI bias.

Wu~et~al.~\cite{wu2022quantifying} proposes the allocation-deterioration framework for detecting and quantifying health inequalities induced by AI models. This framework quantifies inequalities as the area between two allocation-deterioration curves. They conducted experiments on synthetic datasets and real-world ICU datasets to assess the framework's performance and applied the framework to the ICU dataset and quantified the unfairness of AI algorithms between White and Non-White patients. So et al.~\cite{so2022beyond} explores the limitations of fairness in machine learning and proposes a reparative approach to address historical housing discrimination in the US. In that work, they used contemporary mortgage data and historical census data to conduct case studies to demonstrate the impact of historical discrimination on wealth accumulation and estimate housing compensation costs. They then proposed a remediation framework that includes analysing historical biases, intervening in algorithmic systems, and developing machine learning processes that reduce correct historical harms.

\textbf{Latent Class Analysis (LCA)} is a statistical method based on mixture models and often used to detect potential or unobserved heterogeneity in samples \cite{hagenaars2002applied}. By analysing response patterns of observed variables, LCA can identify potential subgroups within a sample set \cite{muthen2000integrating}. The basic idea of LCA is that some parameters of a postulated statistical model differ across unobserved subgroups, forming the categories of a categorical latent variable \cite{vermunt2004latent}. In 1950, Lazarsfeld \cite{lazarsfeld1950logical} introduced LCA as a means of constructing typologies or clusters using dichotomous observed variables. Over two decades later, Goodman \cite{goodman1974analysis} enhanced the model's practical applicability by devising an algorithm for obtaining maximum likelihood estimates of its parameters. Since then, many new frameworks have been proposed, including models with continuous covariates, local dependencies, ordinal variables, multiple latent variables, and repeated measures \cite{vermunt2004latent}.

Because LCA is a person-centered mixture model, it is widely used in sociology and statistics to interpret and identify different subgroups in a population that often share certain external characteristics from data \cite{weller2020latent}. However, in social sciences, LCA is used in cross-sectional and longitudinal studies. For example, in relevant studies in psychology \cite{martinez2020cyberbullying}, social sciences \cite{bardazzi2023energy}, and epidemiology \cite{sinha2021latent}, mixed models and LCA can be used to establish probabilistic diagnoses when no suitable gold standard is available \cite{morin2023stepmix}. 

In \cite{martinez2020cyberbullying}, the relationship between cyberbullying and social anxiety among Hispanic adolescents was explored. The sample consisted of 1,412 Spanish secondary school students aged 12 to 18 years. There were significant differences in cyberbullying patterns across all social anxiety subscales after applying LCA. Compared with other profiles, students with higher cyberbullying traits scored higher on social avoidance and distress in social situations, as well as lower levels of fear of negative evaluation and distress in new situations. Researchers in \cite{bardazzi2023energy} developed a tool, using LCA, to characterise energy poverty without the need to arbitrarily define binary cutoffs. The authors highlight the need for a multidimensional approach to measuring energy poverty and discuss the challenges of identifying vulnerable consumers. The research in \cite{sinha2021latent} aimed to identify subgroups in COVID-19-related acute respiratory distress syndrome (ARDS) and compare them with previously described ARDS subphenotypes by using LCA. The study found that there were two COVID-19-related ARDS subgroups with differential outcomes, similar to previously described ARDS subphenotypes.

\section{Quantifying the Cross-sectoral Intersecting Discrepancies}
\label{sec:method}

\begin{figure}
    \centering
    \includegraphics[width=0.5\textwidth]{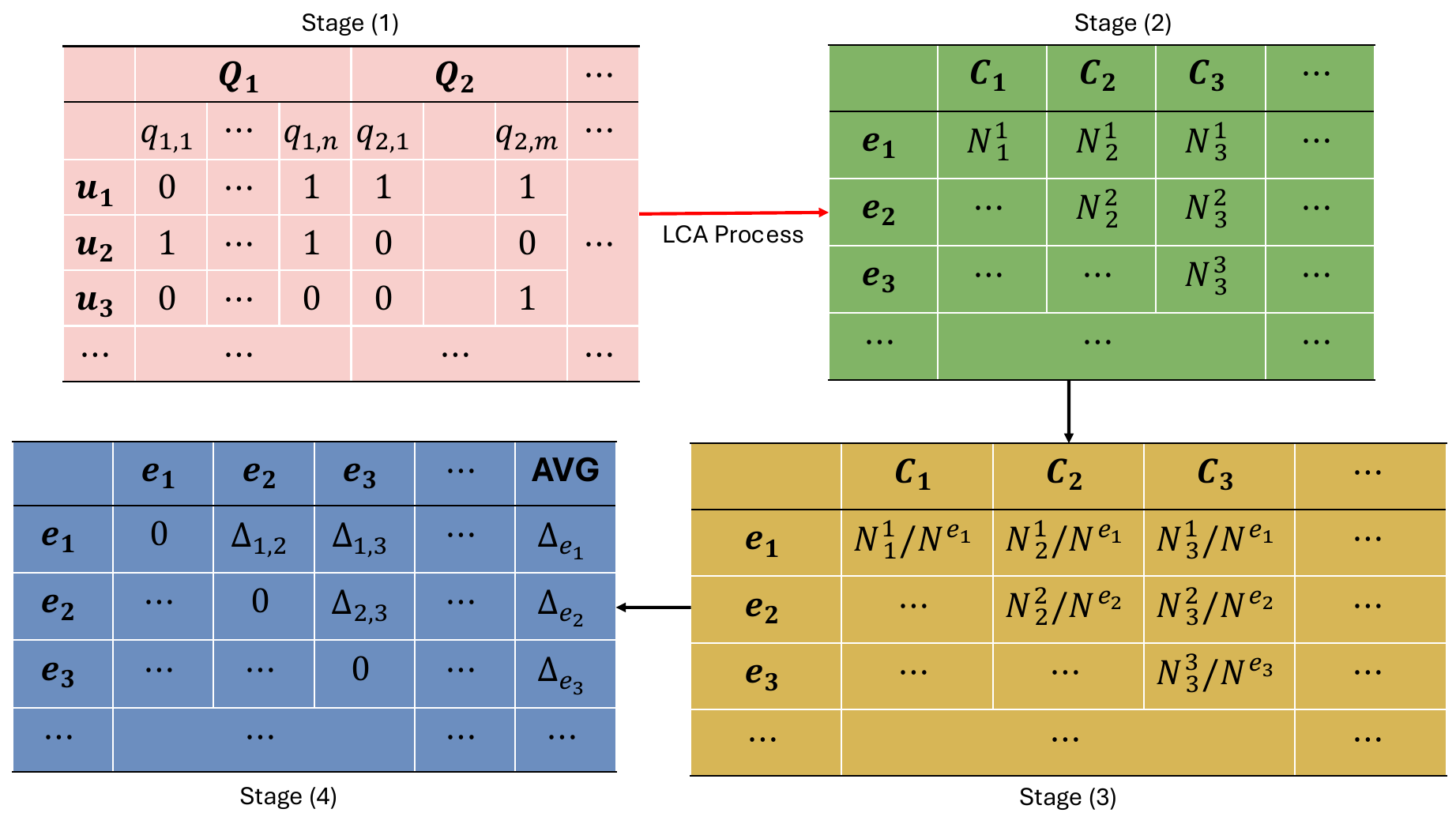}
    \caption{The process of the proposed approach for quantifying cross-sectoral discrepancies within different groups.}
    \label{fig:workflow}
\end{figure}

The overall workflow of the proposed approach is shown in \Cref{fig:workflow}, using a binary-encoded survey data as an example. It is noted that our approach is not limited to this specific format and can be applied to a wide range of similar problems. We will present more experiments with other datasets in \Cref{sec:exp} to further validate and showcase the features of our approach.

In \Cref{fig:workflow}, Stage (1) illustrates the binary-encoded data \( D \), where \( \{Q_1, Q_2, \dots\} \) represents the selected survey questions, covering user experiences across different sectors. Similarly, \( \{u_1, u_2, u_3, \dots\} \) denotes the set of survey respondents. Here, \( q \) represents the response options; for example, \( q_{2,1} \) refers to the first option for \( Q_2 \). A value of `1' indicates a selected option, while `0' signifies that it was not selected. Each respondent's responses can be represented as a vector \( \mathbf{q} \). The set of all indicator variables is denoted by \( X \), with \( \mathbf{q} \in \mathcal{X} \), where $\mathcal{X}$ is the space of all possible response vectors. For other datasets, the encoding method should be selected based on the data format and type. The red arrow in \Cref{fig:workflow} illustrates the LCA process\footnote{StepMix (\url{https://stepmix.readthedocs.io/en/latest/index.html}) Python repository is used to implement LCA in this research.}, which includes hyperparameter selection and model fitting.

For simplicity, the LCA process is defined in \Cref{eq:lca}, where \( \theta_C \) and \( \theta_X \) specify the marginal distribution of the latent classes \( C \) and the class-conditional distribution of the indicator variables $X$, respectively. Here, \( \theta = \left(\theta_C, \theta_X\right) \), and each latent class is denoted by \( c \). In fact, the LCA is specified by a set of $\theta \in \Theta$, where $\left(\theta_C, \theta_X\right) \in \left(\Theta_C \times \Theta_X\right)$.

\begin{equation}
\label{eq:lca}
   p(c, \mathbf{q} ; \theta)=p\left(c ; \theta_C\right) p\left(\mathbf{q} \mid c ; \theta_X\right)
\end{equation}

The advantage of using LCA is straightforward: it considers the joint probability distribution of all variables. This means potential inequalities or discrepancies can be analysed jointly. Once we obtain the distributions of latent classes $\{c_1, c_2, c_3, ...\}$ over user-defined groups $\{e_1, e_2, e_3, ...\}$ (as shown in \Cref{fig:workflow} Stage (2)), we can calculate the discrepancy $\Delta$.

\begin{algorithm}\footnotesize
\caption{\footnotesize{Quantifying the Intersecting Discrepancies within Multiple Groups}}\label{alg:qd}
\begin{algorithmic}[1]
\State \text {Input: } $D$ and $G$ $ \triangleright \text { G denotes a set of user-defined groups}$
\State \text {Initialise } $M$ $ \quad\quad\triangleright \text { Create LCA model $M$}$
\State \text{Estimate} $M$ based on $D$
\For{$e$ \textbf{in} $G$}
\For{$c$ \textbf{in} $C$}
            \State $r_c^e=N_c^e / N^e$
\EndFor
\EndFor
\For{$e$ \textbf{in} $G$}
\For{$e^\prime$ \textbf{in} $G$}
    \State $\Delta_{ee^\prime} = 1- \frac{\mathbf{r}_e \cdot \mathbf{r}_{e^{\prime}}}{\left\|\mathbf{r}_e\right\|\left\|\mathbf{r}_{e^{\prime}}\right\|}$ $\quad\triangleright \text {Pair-wise Calculation}$
\EndFor
\EndFor
\State \text {Output: } Discrepancy matrix $S$ of size $|G|\times|G|$

\end{algorithmic}
\end{algorithm}
\textbf{Quantification of Discrepancy} Let us denote the size of the dataset as $N$, where $i \in {1, 2, \dots, N}$ represents an individual sample. Concurrently, let $c \in C$ denote a latent class, with the total number of classes being $|C|$, and let $N_c$ represent the count of samples classified into latent class $c$. To quantify the discrepancies, it is necessary to establish a grouping variable $G$, which can be defined based on factors such as ethnicity, age, or income level. Here, $|G|$ denotes the total number of user-defined groups, $e \in G$ represents one specific group within this set, $N^e$ denotes the number of individuals from group $e$, and $N^e_c$ denotes the number of individuals from group $e$ assigned to the class $c$.

To initiate the quantification process, the proportions $r$ of samples from each user-defined group within each latent class need to be calculated, as detailed in Line 6 in \Cref{alg:qd}.  This calculation can be performed using $r_c^e = N^{e}_{c}/ N^{e}, \forall e \in G, c \in C$. The reason for calculating $r$ is that user-defined groups may have different numbers of samples; therefore, using percentages for subsequent analyses ensures fairness and consistency.

Subsequently, we can derive a matrix of results characterised by dimensions $|G| \times |C|$, as shown in \Cref{fig:workflow}~(3). Within this matrix, each row corresponds to the proportions of samples from a specific group within each latent class. It is important to note that in this context, each latent class effectively represents an individual user profile and can be viewed as a distinctive feature. Consequently, each row within the matrix may be employed as a feature vector denoted as $\mathbf{r}_{e}$, serving as a representation of a specific group within the feature space.

In the assessment of discrepancy between two feature vectors, various methods may be employed, including the Euclidean distance, Kullback-Leibler Divergence, Earth Mover's Distance, and Manhattan Distance, among others. In our approach, we propose the utilisation of Cosine Similarity to calculate the discrepancy, which is defined as $ \Delta= 1- \cos (\theta)= 1- \frac{\mathbf{r}_{e} \cdot \mathbf{r}_{{e}^{\prime}}}{\|\mathbf{r}_{e}\|\|\mathbf{r}_{e^{\prime}}\|}$. Finally, we can iteratively calculate $\Delta$ between any pairs of vectors $\mathbf{r}$ and obtain the discrepancy matrix $S$ (as shown in Stage (4) of \Cref{fig:workflow}). The AVG column in Stage (4) contains the mean discrepancy values for $e$, which can be viewed an approximation for how each $e$ is different from others. 

We suggest the use of Cosine Similarity due to its inherent characteristics, including a natural value range spanning from 0 to 1 as $\mathbf{r}_{e}$ and $\mathbf{r}_{e^{\prime}}$ contain no negative values. Importantly, it does not necessitate additional normalisation procedures. This metric, possessing with a fixed value range, enhances comparability and offers support to subsequent AI fairness research. The proposed approach is summarised in Algorithm \ref{alg:qd}.

\section{Experiments}
\label{sec:exp}
\subsection{The \PRIMEproject}
Within the \textit{\PRIME} project\footnote{ More dataset details can be found in the Appendix \ref{appendix:prime}.}, a systematic online survey was conducted to explore the experiences of individuals from ME groups in the UK, focusing on digitalised health, energy, and housing aspects. To examine cross-sectoral intersecting discrepancies among the seven ethnic groups (as shown in \Cref{tab:primediscrepancy}), we selected three questions from the \textit{\PRIME}survey data related to these sectors. The answers from ME participants formed the dataset for this experiment. For England and Scotland, we have 594 and 284 samples, respectively. Due to varying sample sizes across ethnicities, we calculated discrepancies separately for each region.

Selecting the number of latent classes in LCA requires presetting. To address this, we conducted hyperparameter optimisation for each experiment to find the elbow point, applying this optimisation to all subsequent experiments.

The discrepancy results for England are presented in the left part of \Cref{tab:primediscrepancy}. The table shows that the Chinese group has the largest average (AVG) discrepancy value compared to other groups. Meanwhile, the Indian group exhibits the smallest discrepancy with the Bangladeshi group and also shows similarity to the Pakistani group. This is likely due to their close geographical locations and similar cultural backgrounds and lifestyles. 

The right part of \Cref{tab:primediscrepancy} presents the results for Scotland, showing similar outcomes: the Chinese group is distinct from others, while the Bangladeshi group is similar to the Pakistani group. The differences between England and Scotland may be attributed to their different policies and circumstances. We hypothesise that the primary reason for the Chinese group standing out is a lack of English proficiency. This is supported by our preliminary research mentioned in \Cref{sec:introduction}, which shows a significant number of Chinese participants expressing this concern. Additionally, BBC News \cite{bbc2024} reports that the Chinese community experiences some of the highest rates of racism among all ethnic groups in the UK. Our discrepancy values may help explain this, as the Chinese group shows different experiences in digitalised online services.

\begin{figure}
    \centering
    \includegraphics[width=0.6\linewidth]{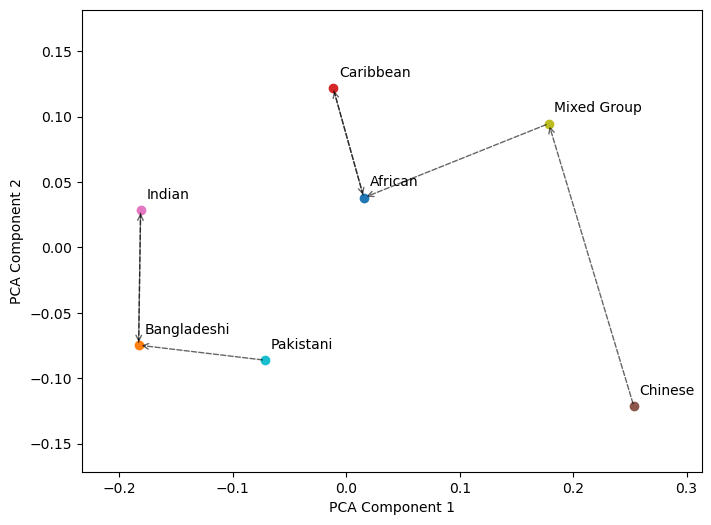}
    \caption{The PCA visualisation of relationships among ME groups based on percentage distributions across latent classes on PRIME dataset (England). The dashed line with an arrow indicates the distance from one dot to its nearest neighbour.}
    \label{fig:pca}
\end{figure}

To further verify the reliability of the discrepancy computation, we used PCA dimensionality reduction to visualise the relationships among ME groups based on the percentage distributions (as shown in \Cref{fig:workflow} (3)) of each ME group across different latent classes. The patterns presented in \Cref{fig:pca} align with the findings obtained from the analysis of \Cref{tab:primediscrepancy} (England). Overall, the Chinese group continues to stand out, positioned further away from other groups, indicating it has the highest average discrepancy. The Mixed and Bangladeshi groups are relatively distant from others, as they are scattered on opposite sides. Indian, Bangladeshi, and Pakistani groups are closer to each other compared to other groups, implying a similarity supported by the discrepancy values in \Cref{tab:primediscrepancy}. Meanwhile, the African and Caribbean groups are the nearest to each other, as represented by the double arrows between them.

\begin{table*}
\caption{The matrix of discrepancies between 7 ethnic groups for \textit{\PRIME}project's England and Scotland data. The AVG denotes the average discrepancy value for one group.}
\label{tab:primediscrepancy}
\centering
\setlength{\extrarowheight}{0pt}
\addtolength{\extrarowheight}{\aboverulesep}
\addtolength{\extrarowheight}{\belowrulesep}
\setlength{\aboverulesep}{0pt}
\setlength{\belowrulesep}{0pt}
\resizebox{0.8\textwidth}{!}{
\begin{tabular}{ccccccccc!{\vrule width \lightrulewidth}cccccccc} 
\toprule
\multicolumn{1}{l}{} & \multicolumn{7}{c}{\textbf{England}}                                                                                                                                                                                                                                                                                     & \multicolumn{1}{l!{\vrule width \lightrulewidth}}{} & \multicolumn{8}{c}{\textbf{Scotland}}                                                                                                                                                                                                                                                                                                                                  \\ 
\cmidrule(lr){2-17}
\multicolumn{1}{l}{} & \textbf{African}                           & \textbf{Bangladeshi}                       & \textbf{Caribbean}                         & \textbf{Chinese}                           & \textbf{Indian}                            & \textbf{Mixed Group}                       & \textbf{Pakistani}                         & \textbf{AVG}                                        & \textbf{African}                           & \textbf{Bangladeshi}                       & \textbf{Caribbean}                         & \textbf{Chinese}                           & \textbf{Indian}                            & \textbf{Mixed Group}                       & \textbf{Pakistani}                         & \textbf{AVG}                                \\
\textbf{African}     & {\cellcolor[rgb]{0.973,0.412,0.42}}0.0000  & {\cellcolor[rgb]{0.973,0.549,0.557}}0.0899 & {\cellcolor[rgb]{0.973,0.471,0.478}}0.0383 & {\cellcolor[rgb]{0.976,0.69,0.698}}0.1810  & {\cellcolor[rgb]{0.973,0.494,0.502}}0.0547 & {\cellcolor[rgb]{0.973,0.502,0.51}}0.0590  & {\cellcolor[rgb]{0.973,0.49,0.498}}0.0517  & {\cellcolor[rgb]{0.388,0.745,0.482}}0.0678          & {\cellcolor[rgb]{0.973,0.412,0.42}}0.0000  & {\cellcolor[rgb]{0.973,0.514,0.522}}0.0666 & {\cellcolor[rgb]{0.973,0.455,0.463}}0.0296 & {\cellcolor[rgb]{0.98,0.714,0.722}}0.1956  & {\cellcolor[rgb]{0.973,0.463,0.471}}0.0342 & {\cellcolor[rgb]{0.973,0.427,0.435}}0.0118 & {\cellcolor[rgb]{0.973,0.443,0.455}}0.0227 & {\cellcolor[rgb]{0.388,0.745,0.482}}0.0515  \\
\textbf{Bangladeshi} & {\cellcolor[rgb]{0.973,0.549,0.557}}0.0899 & {\cellcolor[rgb]{0.973,0.412,0.42}}0.0000  & {\cellcolor[rgb]{0.976,0.62,0.627}}0.1359  & {\cellcolor[rgb]{0.988,0.988,1}}0.3734     & {\cellcolor[rgb]{0.973,0.439,0.447}}0.0200 & {\cellcolor[rgb]{0.98,0.831,0.843}}0.2738  & {\cellcolor[rgb]{0.973,0.459,0.467}}0.0308 & {\cellcolor[rgb]{0.639,0.847,0.698}}0.1320          & {\cellcolor[rgb]{0.973,0.514,0.522}}0.0666 & {\cellcolor[rgb]{0.973,0.412,0.42}}0.0000  & {\cellcolor[rgb]{0.973,0.459,0.467}}0.0324 & {\cellcolor[rgb]{0.984,0.882,0.894}}0.3043 & {\cellcolor[rgb]{0.976,0.565,0.573}}0.0989 & {\cellcolor[rgb]{0.973,0.498,0.506}}0.0563 & {\cellcolor[rgb]{0.973,0.427,0.435}}0.0118 & {\cellcolor[rgb]{0.49,0.784,0.569}}0.0815   \\
\textbf{Caribbean}   & {\cellcolor[rgb]{0.973,0.471,0.478}}0.0383 & {\cellcolor[rgb]{0.976,0.62,0.627}}0.1359  & {\cellcolor[rgb]{0.973,0.412,0.42}}0.0000  & {\cellcolor[rgb]{0.98,0.788,0.8}}0.2456    & {\cellcolor[rgb]{0.973,0.529,0.537}}0.0764 & {\cellcolor[rgb]{0.976,0.584,0.592}}0.1131 & {\cellcolor[rgb]{0.976,0.557,0.565}}0.0951 & {\cellcolor[rgb]{0.518,0.796,0.592}}0.1006          & {\cellcolor[rgb]{0.973,0.455,0.463}}0.0296 & {\cellcolor[rgb]{0.973,0.459,0.467}}0.0324 & {\cellcolor[rgb]{0.973,0.412,0.42}}0.0000  & {\cellcolor[rgb]{0.984,0.941,0.953}}0.3430 & {\cellcolor[rgb]{0.973,0.439,0.447}}0.0191 & {\cellcolor[rgb]{0.973,0.494,0.502}}0.0546 & {\cellcolor[rgb]{0.973,0.435,0.443}}0.0159 & {\cellcolor[rgb]{0.451,0.769,0.537}}0.0706  \\
\textbf{Chinese}     & {\cellcolor[rgb]{0.976,0.69,0.698}}0.1810  & {\cellcolor[rgb]{0.988,0.988,1}}0.3734     & {\cellcolor[rgb]{0.98,0.788,0.8}}0.2456    & {\cellcolor[rgb]{0.973,0.412,0.42}}0.0000  & {\cellcolor[rgb]{0.984,0.98,0.992}}0.3700  & {\cellcolor[rgb]{0.976,0.596,0.604}}0.1201 & {\cellcolor[rgb]{0.98,0.788,0.8}}0.2459    & {\cellcolor[rgb]{0.988,0.988,1}}0.2194              & {\cellcolor[rgb]{0.98,0.714,0.722}}0.1956  & {\cellcolor[rgb]{0.984,0.882,0.894}}0.3043 & {\cellcolor[rgb]{0.984,0.941,0.953}}0.3430 & {\cellcolor[rgb]{0.973,0.412,0.42}}0.0000  & {\cellcolor[rgb]{0.988,0.988,1}}0.3717     & {\cellcolor[rgb]{0.976,0.616,0.627}}0.1334 & {\cellcolor[rgb]{0.98,0.788,0.8}}0.2438    & {\cellcolor[rgb]{0.988,0.988,1}}0.2274      \\
\textbf{Indian}      & {\cellcolor[rgb]{0.973,0.494,0.502}}0.0547 & {\cellcolor[rgb]{0.973,0.439,0.447}}0.0200 & {\cellcolor[rgb]{0.973,0.529,0.537}}0.0764 & {\cellcolor[rgb]{0.984,0.98,0.992}}0.3700  & {\cellcolor[rgb]{0.973,0.412,0.42}}0.0000  & {\cellcolor[rgb]{0.98,0.741,0.749}}0.2139  & {\cellcolor[rgb]{0.973,0.459,0.467}}0.0311 & {\cellcolor[rgb]{0.553,0.812,0.624}}0.1094          & {\cellcolor[rgb]{0.973,0.463,0.471}}0.0342 & {\cellcolor[rgb]{0.976,0.565,0.573}}0.0989 & {\cellcolor[rgb]{0.973,0.439,0.447}}0.0191 & {\cellcolor[rgb]{0.988,0.988,1}}0.3717     & {\cellcolor[rgb]{0.973,0.412,0.42}}0.0000  & {\cellcolor[rgb]{0.973,0.537,0.545}}0.0821 & {\cellcolor[rgb]{0.973,0.498,0.506}}0.0575 & {\cellcolor[rgb]{0.533,0.804,0.608}}0.0948  \\
\textbf{Mixed Group} & {\cellcolor[rgb]{0.973,0.502,0.51}}0.0590  & {\cellcolor[rgb]{0.98,0.831,0.843}}0.2738  & {\cellcolor[rgb]{0.976,0.584,0.592}}0.1131 & {\cellcolor[rgb]{0.976,0.596,0.604}}0.1201 & {\cellcolor[rgb]{0.98,0.741,0.749}}0.2139  & {\cellcolor[rgb]{0.973,0.412,0.42}}0.0000  & {\cellcolor[rgb]{0.98,0.718,0.725}}0.1987  & {\cellcolor[rgb]{0.671,0.859,0.725}}0.1398          & {\cellcolor[rgb]{0.973,0.427,0.435}}0.0118 & {\cellcolor[rgb]{0.973,0.498,0.506}}0.0563 & {\cellcolor[rgb]{0.973,0.494,0.502}}0.0546 & {\cellcolor[rgb]{0.976,0.616,0.627}}0.1334 & {\cellcolor[rgb]{0.973,0.537,0.545}}0.0821 & {\cellcolor[rgb]{0.973,0.412,0.42}}0.0000  & {\cellcolor[rgb]{0.973,0.443,0.451}}0.0209 & {\cellcolor[rgb]{0.388,0.745,0.482}}0.0513  \\
\textbf{Pakistani}   & {\cellcolor[rgb]{0.973,0.49,0.498}}0.0517  & {\cellcolor[rgb]{0.973,0.459,0.467}}0.0308 & {\cellcolor[rgb]{0.976,0.557,0.565}}0.0951 & {\cellcolor[rgb]{0.98,0.788,0.8}}0.2459    & {\cellcolor[rgb]{0.973,0.459,0.467}}0.0311 & {\cellcolor[rgb]{0.98,0.718,0.725}}0.1987  & {\cellcolor[rgb]{0.973,0.412,0.42}}0.0000  & {\cellcolor[rgb]{0.486,0.784,0.569}}0.0933          & {\cellcolor[rgb]{0.973,0.443,0.455}}0.0227 & {\cellcolor[rgb]{0.973,0.427,0.435}}0.0118 & {\cellcolor[rgb]{0.973,0.435,0.443}}0.0159 & {\cellcolor[rgb]{0.98,0.788,0.8}}0.2438    & {\cellcolor[rgb]{0.973,0.498,0.506}}0.0575 & {\cellcolor[rgb]{0.973,0.443,0.451}}0.0209 & {\cellcolor[rgb]{0.973,0.412,0.42}}0.0000  & {\cellcolor[rgb]{0.392,0.745,0.486}}0.0532  \\
\bottomrule
\end{tabular}}
\end{table*}

\subsection{EVENS}
The Centre on the Dynamics of Ethnicity (CoDE), funded by the Economic and Social Research Council (ESRC), conducted ``The COVID Race Inequalities Programme''. As part of this project, CoDE carried out the Evidence for Equality National Survey (EVENS)\footnote{\url{https://www.evensurvey.co.uk/}}, which documents the lives of ethnic and religious minorities in Britain during the coronavirus pandemic. The EVENS dataset comprises 14,215 data points and categorizes participants into 18 different ethnic minority groups. To facilitate a comparative analysis with the \textit{\PRIME}project's experiment, we focused on the same seven ethnic groups from \textit{\PRIME}project, resulting in a filtered dataset of 4,348 participants from England and 253 participants from Scotland. We excluded entries with missing values to ensure the robustness of our analysis.

It is important to note that the EVENS dataset uses a different definition for mixed or multiple ethnic groups compared to the \textit{\PRIME}project. To avoid inconsistencies that could affect the final analysis, we excluded mixed or multiple ethnic groups from the EVENS calculations. Due to the different questions in the EVENS and \textit{\PRIME}project surveys, we selected three types of cross-sectoral questions from EVENS for analysis: housing, experiences of harassment, and financial situation. These questions were chosen to provide a broad view of participants' living conditions, social experiences, and economic status during the pandemic.

We applied LCA separately to the data from England and Scotland to uncover patterns and discrepancies within and between these regions. This approach allows us to identify distinct subgroups within the ethnic communities based on their responses to the selected questions, providing deeper insights into the intersecting cross-sectoral experiences of these groups.

Seeing the discrepancies in the EVENS England data, shown in the left part of \Cref{tab:evensdiscrepancies}, it is clear that all discrepancy values are relatively small compared to the \textit{\PRIME}project results. Notably, there are large discrepancies between the Indian and Bangladeshi groups, and between the Indian and Chinese groups. This differs slightly from the conclusions of the \textit{\PRIME}project's experiment, likely due to the different types of cross-sectoral questions selected. In the AVG column, the Bangladeshi group has the highest values, indicating they experienced COVID-19 differently. The Census 2021 reported that COVID-19 mortality rates were highest for the Bangladeshi group, for both males and females \cite{Drummond_Pratt_2023}, supporting our findings. The right part of \Cref{tab:evensdiscrepancies} describes the discrepancies for the EVENS Scotland data. There are significant discrepancies between the Caribbean and other ethnic groups, likely related to the unique background of the Caribbean group.

\begin{table*}
\caption{The matrix of discrepancies for EVENS England and Scotland data.}
\label{tab:evensdiscrepancies}
\centering
\setlength{\extrarowheight}{0pt}
\addtolength{\extrarowheight}{\aboverulesep}
\addtolength{\extrarowheight}{\belowrulesep}
\setlength{\aboverulesep}{0pt}
\setlength{\belowrulesep}{0pt}
\resizebox{0.8\textwidth}{!}{
\begin{tabular}{cccccccc|ccccccc} 
\toprule
\multicolumn{1}{l}{} & \multicolumn{7}{c|}{\textbf{England}}                                                                                                                                                                                                                                                                                    & \multicolumn{7}{c}{\textbf{Scotland}}                                                                                                                                                                                                                                                                                     \\ 
\cmidrule(r){2-15}
                     & \textbf{African}                           & \textbf{Bangladeshi}                       & \textbf{Caribbean}                         & \textbf{Chinese}                           & \textbf{Indian}                            & \textbf{Pakistani}                         & \textbf{AVG}                               & \textbf{African}                           & \textbf{Bangladeshi}                       & \textbf{Caribbean}                         & \textbf{Chinese}                           & \textbf{Indian}                            & \textbf{Pakistani}                         & \textbf{AVG}                                \\
\textbf{African}     & {\cellcolor[rgb]{0.973,0.412,0.42}}0.0000  & {\cellcolor[rgb]{0.976,0.694,0.702}}0.0112 & {\cellcolor[rgb]{0.973,0.443,0.451}}0.0014 & {\cellcolor[rgb]{0.973,0.506,0.514}}0.0038 & {\cellcolor[rgb]{0.976,0.569,0.576}}0.0062 & {\cellcolor[rgb]{0.973,0.455,0.463}}0.0018 & {\cellcolor[rgb]{0.467,0.776,0.549}}0.0040 & {\cellcolor[rgb]{0.973,0.412,0.42}}0.0000  & {\cellcolor[rgb]{0.973,0.435,0.443}}0.0246 & {\cellcolor[rgb]{0.988,0.988,1}}0.5408     & {\cellcolor[rgb]{0.973,0.443,0.451}}0.0300 & {\cellcolor[rgb]{0.973,0.478,0.486}}0.0637 & {\cellcolor[rgb]{0.976,0.6,0.612}}0.1800   & {\cellcolor[rgb]{0.463,0.773,0.545}}0.1398  \\
\textbf{Bangladeshi} & {\cellcolor[rgb]{0.976,0.694,0.702}}0.0112 & {\cellcolor[rgb]{0.973,0.412,0.42}}0.0000  & {\cellcolor[rgb]{0.976,0.671,0.678}}0.0102 & {\cellcolor[rgb]{0.973,0.49,0.498}}0.0031  & {\cellcolor[rgb]{0.988,0.988,1}}0.0227     & {\cellcolor[rgb]{0.976,0.639,0.647}}0.0090 & {\cellcolor[rgb]{0.988,0.988,1}}0.0094     & {\cellcolor[rgb]{0.973,0.435,0.443}}0.0246 & {\cellcolor[rgb]{0.973,0.412,0.42}}0.0000  & {\cellcolor[rgb]{0.984,0.973,0.984}}0.5283 & {\cellcolor[rgb]{0.973,0.467,0.475}}0.0522 & {\cellcolor[rgb]{0.976,0.573,0.584}}0.1539 & {\cellcolor[rgb]{0.976,0.698,0.706}}0.2697 & {\cellcolor[rgb]{0.537,0.804,0.612}}0.1714  \\
\textbf{Caribbean}   & {\cellcolor[rgb]{0.973,0.443,0.451}}0.0014 & {\cellcolor[rgb]{0.976,0.671,0.678}}0.0102 & {\cellcolor[rgb]{0.973,0.412,0.42}}0.0000  & {\cellcolor[rgb]{0.973,0.529,0.537}}0.0047 & {\cellcolor[rgb]{0.973,0.486,0.494}}0.0030 & {\cellcolor[rgb]{0.973,0.416,0.424}}0.0002 & {\cellcolor[rgb]{0.388,0.745,0.482}}0.0032 & {\cellcolor[rgb]{0.988,0.988,1}}0.5408     & {\cellcolor[rgb]{0.984,0.973,0.984}}0.5283 & {\cellcolor[rgb]{0.973,0.412,0.42}}0.0000  & {\cellcolor[rgb]{0.98,0.831,0.839}}0.3946  & {\cellcolor[rgb]{0.984,0.89,0.902}}0.4505  & {\cellcolor[rgb]{0.976,0.678,0.686}}0.2506 & {\cellcolor[rgb]{0.988,0.988,1}}0.3608      \\
\textbf{Chinese}     & {\cellcolor[rgb]{0.973,0.506,0.514}}0.0038 & {\cellcolor[rgb]{0.973,0.49,0.498}}0.0031  & {\cellcolor[rgb]{0.973,0.529,0.537}}0.0047 & {\cellcolor[rgb]{0.973,0.412,0.42}}0.0000  & {\cellcolor[rgb]{0.98,0.761,0.773}}0.0138  & {\cellcolor[rgb]{0.973,0.529,0.541}}0.0048 & {\cellcolor[rgb]{0.561,0.816,0.631}}0.0050 & {\cellcolor[rgb]{0.973,0.443,0.451}}0.0300 & {\cellcolor[rgb]{0.973,0.467,0.475}}0.0522 & {\cellcolor[rgb]{0.98,0.831,0.839}}0.3946  & {\cellcolor[rgb]{0.973,0.412,0.42}}0.0000  & {\cellcolor[rgb]{0.973,0.478,0.49}}0.0662  & {\cellcolor[rgb]{0.973,0.522,0.529}}0.1036 & {\cellcolor[rgb]{0.388,0.745,0.482}}0.1078  \\
\textbf{Indian}      & {\cellcolor[rgb]{0.976,0.569,0.576}}0.0062 & {\cellcolor[rgb]{0.988,0.988,1}}0.0227     & {\cellcolor[rgb]{0.973,0.486,0.494}}0.0030 & {\cellcolor[rgb]{0.98,0.761,0.773}}0.0138  & {\cellcolor[rgb]{0.973,0.412,0.42}}0.0000  & {\cellcolor[rgb]{0.973,0.514,0.522}}0.0040 & {\cellcolor[rgb]{0.882,0.945,0.906}}0.0083 & {\cellcolor[rgb]{0.973,0.478,0.486}}0.0637 & {\cellcolor[rgb]{0.976,0.573,0.584}}0.1539 & {\cellcolor[rgb]{0.984,0.89,0.902}}0.4505  & {\cellcolor[rgb]{0.973,0.478,0.49}}0.0662  & {\cellcolor[rgb]{0.973,0.412,0.42}}0.0000  & {\cellcolor[rgb]{0.973,0.475,0.482}}0.0589 & {\cellcolor[rgb]{0.443,0.765,0.529}}0.1322  \\
\textbf{Pakistani}   & {\cellcolor[rgb]{0.973,0.455,0.463}}0.0018 & {\cellcolor[rgb]{0.976,0.639,0.647}}0.0090 & {\cellcolor[rgb]{0.973,0.416,0.424}}0.0002 & {\cellcolor[rgb]{0.973,0.529,0.541}}0.0048 & {\cellcolor[rgb]{0.973,0.514,0.522}}0.0040 & {\cellcolor[rgb]{0.973,0.412,0.42}}0.0000  & {\cellcolor[rgb]{0.392,0.745,0.486}}0.0033 & {\cellcolor[rgb]{0.976,0.6,0.612}}0.1800   & {\cellcolor[rgb]{0.976,0.698,0.706}}0.2697 & {\cellcolor[rgb]{0.976,0.678,0.686}}0.2506 & {\cellcolor[rgb]{0.973,0.522,0.529}}0.1036 & {\cellcolor[rgb]{0.973,0.475,0.482}}0.0589 & {\cellcolor[rgb]{0.973,0.412,0.42}}0.0000  & {\cellcolor[rgb]{0.471,0.776,0.553}}0.1438  \\
\midrule
\end{tabular}}
\end{table*}


\subsection{Census 2021 (England and Wales)}
\label{sec: census2021correlation}

\begin{table*}
\centering
\caption{The matrices for the discrepancies of Census 2021 and the deprivation discrepancies between LSOAs across user-defined ME population percentage groups based on Deprivation 2019.}
\label{tab:censusdiscrepancies}
\setlength{\extrarowheight}{0pt}
\addtolength{\extrarowheight}{\aboverulesep}
\addtolength{\extrarowheight}{\belowrulesep}
\setlength{\aboverulesep}{0pt}
\setlength{\belowrulesep}{0pt}
\resizebox{0.6\textwidth}{!}{
\begin{tabular}{ccccccc|cccccc} 
\toprule
                  & \multicolumn{6}{c|}{\textbf{Census~}}                                                                                                                                                                                                                                       & \multicolumn{6}{c}{\textbf{Deprivation}}                                                                                                                                                                                                                                     \\ 
\cmidrule(r){2-13}
                  & \textbf{0-20\%}                            & \textbf{20-40\%}                           & \textbf{40-60\%}                           & \textbf{60-80\%}                           & \textbf{80-100\%}                          & \textbf{AVG}                               & \textbf{0-20\%}                            & \textbf{20-40\%}                           & \textbf{40-60\%}                           & \textbf{60-80\%}                           & \textbf{80-100\%}                          & \textbf{AVG}                                \\
\textbf{0-20\%}   & {\cellcolor[rgb]{0.973,0.412,0.42}}0.0000  & {\cellcolor[rgb]{0.98,0.71,0.722}}0.4865   & {\cellcolor[rgb]{0.98,0.827,0.839}}0.6783  & {\cellcolor[rgb]{0.984,0.941,0.953}}0.8603 & {\cellcolor[rgb]{0.988,0.988,1}}0.9347     & {\cellcolor[rgb]{0.988,0.988,1}}0.5920     & {\cellcolor[rgb]{0.973,0.412,0.42}}0.0000  & {\cellcolor[rgb]{0.976,0.639,0.647}}0.2001 & {\cellcolor[rgb]{0.98,0.741,0.749}}0.2896  & {\cellcolor[rgb]{0.984,0.851,0.863}}0.3877 & {\cellcolor[rgb]{0.988,0.988,1}}0.5064     & {\cellcolor[rgb]{0.988,0.988,1}}0.2768      \\
\textbf{20-40\%}  & {\cellcolor[rgb]{0.98,0.71,0.722}}0.4865   & {\cellcolor[rgb]{0.973,0.412,0.42}}0.0000  & {\cellcolor[rgb]{0.973,0.494,0.502}}0.1371 & {\cellcolor[rgb]{0.976,0.651,0.663}}0.3934 & {\cellcolor[rgb]{0.98,0.753,0.765}}0.5565  & {\cellcolor[rgb]{0.51,0.792,0.588}}0.3147  & {\cellcolor[rgb]{0.976,0.639,0.647}}0.2001 & {\cellcolor[rgb]{0.973,0.412,0.42}}0.0000  & {\cellcolor[rgb]{0.973,0.447,0.455}}0.0313 & {\cellcolor[rgb]{0.973,0.537,0.545}}0.1123 & {\cellcolor[rgb]{0.976,0.659,0.671}}0.2203 & {\cellcolor[rgb]{0.459,0.773,0.545}}0.1128  \\
\textbf{40-60\%}  & {\cellcolor[rgb]{0.98,0.827,0.839}}0.6783  & {\cellcolor[rgb]{0.973,0.494,0.502}}0.1371 & {\cellcolor[rgb]{0.973,0.412,0.42}}0.0000  & {\cellcolor[rgb]{0.973,0.482,0.49}}0.1173  & {\cellcolor[rgb]{0.976,0.58,0.588}}0.2744  & {\cellcolor[rgb]{0.388,0.745,0.482}}0.2414 & {\cellcolor[rgb]{0.98,0.741,0.749}}0.2896  & {\cellcolor[rgb]{0.973,0.447,0.455}}0.0313 & {\cellcolor[rgb]{0.973,0.412,0.42}}0.0000  & {\cellcolor[rgb]{0.973,0.447,0.455}}0.0314 & {\cellcolor[rgb]{0.973,0.518,0.529}}0.0963 & {\cellcolor[rgb]{0.388,0.745,0.482}}0.0897  \\
\textbf{60-80\%}  & {\cellcolor[rgb]{0.984,0.941,0.953}}0.8603 & {\cellcolor[rgb]{0.976,0.651,0.663}}0.3934 & {\cellcolor[rgb]{0.973,0.482,0.49}}0.1173  & {\cellcolor[rgb]{0.973,0.412,0.42}}0.0000  & {\cellcolor[rgb]{0.973,0.439,0.447}}0.0445 & {\cellcolor[rgb]{0.459,0.773,0.541}}0.2831 & {\cellcolor[rgb]{0.984,0.851,0.863}}0.3877 & {\cellcolor[rgb]{0.973,0.537,0.545}}0.1123 & {\cellcolor[rgb]{0.973,0.447,0.455}}0.0314 & {\cellcolor[rgb]{0.973,0.412,0.42}}0.0000  & {\cellcolor[rgb]{0.973,0.443,0.451}}0.0283 & {\cellcolor[rgb]{0.459,0.773,0.541}}0.1119  \\
\textbf{80-100\%} & {\cellcolor[rgb]{0.988,0.988,1}}0.9347     & {\cellcolor[rgb]{0.98,0.753,0.765}}0.5565  & {\cellcolor[rgb]{0.976,0.58,0.588}}0.2744  & {\cellcolor[rgb]{0.973,0.439,0.447}}0.0445 & {\cellcolor[rgb]{0.973,0.412,0.42}}0.0000  & {\cellcolor[rgb]{0.592,0.827,0.659}}0.3620 & {\cellcolor[rgb]{0.988,0.988,1}}0.5064     & {\cellcolor[rgb]{0.976,0.659,0.671}}0.2203 & {\cellcolor[rgb]{0.973,0.518,0.529}}0.0963 & {\cellcolor[rgb]{0.973,0.443,0.451}}0.0283 & {\cellcolor[rgb]{0.973,0.412,0.42}}0.0000  & {\cellcolor[rgb]{0.643,0.847,0.702}}0.1703  \\
\bottomrule
\end{tabular}}
\end{table*}

To further test our approach, we applied it to the Census 2021 dataset \cite{Census2021}, which gathers information on individuals and households in England and Wales every decade. These data help plan and finance essential local services. We compared our results with the UK deprivation indices data from 2019 \cite{UKdeprivation}, which classify relative deprivation in small areas. We hypothesised that discrepancy values should correlate with the deprivation indices, reflecting discrepancies across energy, health, housing, and socioeconomic sectors. The key difference is that our discrepancy values are data-driven, while deprivation indices are based on human-centred assessments, suggesting that our approach is complementary.

For our experiments, we selected four cross-sectoral questions from the census related to energy (type of central heating), health (general health), housing (occupancy rating for bedrooms), and socioeconomic status (household deprivation). Note that the socioeconomic data in Census 2021 differ from the 2019 deprivation indices due to different definitions and coverage \cite{Census2021,UKdeprivation}.

Additionally, for Census 2021, we selected Lower Layer Super Output Areas (LSOAs) \cite{Census2021} as samples instead of individuals, as individual data were not accessible. After cleaning the data and removing unmatched LSOAs, we had 31,810 LSOAs in total. Unmatched LSOAs, which appear only in either Census 2021 or Deprivation 2019, were removed. In the Deprivation 2019 dataset, each LSOA is labelled with a deprivation level from 1 to 10 (1 being the most deprived). As our samples are LSOAs, we quantified discrepancies between different LSOAs. Since the raw data does not include group attributes, we classified LSOAs into five groups based on the percentage of the population from ME groups: [0\%, 20\%), [20\%, 40\%), [40\%, 60\%), [60\%, 80\%), and [80\%, 100\%].

The proposed approach quantifies the discrepancies between the defined ME population-related groups based on the selected Census 2021 data. The results are shown in the left part of \Cref{tab:censusdiscrepancies}. It is evident that the discrepancies between LSOAs increase as differences in ME population percentages increase, indicating significant disparities in living conditions for ME individuals across different LSOAs, particularly in terms of energy, housing, and health aspects. Notably, the 0\%-20\% group shows the largest AVG discrepancy compared to other groups, suggesting that White individuals in those LSOAs experience significantly different living conditions. Since the 0\%-20\% group constitutes a large portion of the UK (see \Cref{appendix:census}), this finding suggests potential unequal treatment and possible neglect of other LSOAs. Additionally, the 40\%-60\% group has the smallest AVG discrepancy value, likely due to its intermediate position among the ME groups, sharing characteristics with the 0\%-20\% and 20\%-40\% groups, as well as 60\%-80\% and 80\%-100\% groups.

\textbf{Correlation Analysis} Furthermore, based on the deprivation indices from Deprivation 2019 \cite{UKdeprivation} and the defined groups, we calculated the percentages of LSOAs in each deprivation-labelled group across various ME population groups. The results are shown in Appendix \ref{appendix:census}, and we treated each row as a feature vector representing one group of LSOAs. We then iteratively calculated the deprivation discrepancies for each pair of rows, with the results presented in the right part of \Cref{tab:censusdiscrepancies}. We observed similar patterns (color change) to those in the left part of \Cref{tab:censusdiscrepancies}, which can verify the reliability of our proposed approach.

To statistically verify our proposed approach, we ran Pearson and Spearman row-wise correlation analyses for Census 2021 discrepancies (the left part of \Cref{tab:censusdiscrepancies}) and Deprivation 2019 discrepancies (the right part of \Cref{tab:censusdiscrepancies}). The detailed results are shown in Appendix \ref{appendix:census}. All rows exhibit very strong correlations, implying that our approach can draw conclusions very similar to those of experts. Furthermore, we also flattened both matrices to run a one-time correlation analysis. The Pearson correlation coefficient is 0.9797 with a $p$-value of 1.4437e-17, and the Spearman correlation coefficient is 0.9872 with a $p$-value of 7.436e-20. Both $p$-values are far less than 0.001, indicating a strong correlation.

Additionally, to further assess the performance of LCA, we conducted experiments using $k$-Means as a replacement for LCA to calculate the discrepancies and performed the same correlation analysis. While the correlation coefficients are still strong, they are lower than those achieved by LCA. This demonstrates the efficiency of LCA and further indirectly supports the usability of the proposed framework.

\begin{table}[!htbp]
\centering
\caption{The Discrepancies and False Positive Rate.}
\label{tab: unevenaifairness}
\resizebox{0.45\textwidth}{!}{
\begin{tblr}{
  cells = {c},
  vline{2} = {-}{},
  hline{1,8} = {-}{0.08em},
  hline{2,7} = {-}{},
}
                  & \textbf{0-20\%} & \textbf{20-40\%} & \textbf{40-60\%} & \textbf{60-80\%} & \textbf{80-100\%} \\
\textbf{0-20\%}   & 0.0000          & 0.5021           & 0.6775           & 0.8646           & \textbf{0.9301}   \\
\textbf{20-40\%}  & 0.5021          & 0.0000           & 0.1272           & 0.4158           & 0.5931            \\
\textbf{40-60\%}  & 0.6775          & 0.1272           & 0.0000           & 0.1521           & 0.3217            \\
\textbf{60-80\%}  & 0.8646          & 0.4158           & 0.1521           & 0.0000           & 0.0450            \\
\textbf{80-100\%} & \textbf{0.9301} & 0.5931           & 0.3217           & 0.0450           & 0.0000            \\
\textbf{FPR}      & \textbf{0.1225} & 0.0710           & 0.0421           & 0.0090           & 0                 
\end{tblr}}
\end{table}

\begin{table*}[!htbp]
\caption{Discrepancy Values for Five Groups with Three Sampling Ratios.}
\centering
\label{tab:evensamplingdiscrepancy}
\resizebox{0.9\textwidth}{!}{
\begin{tblr}{
  cells = {c},
  cell{1}{1} = {r=2}{},
  cell{3}{1} = {r=5}{},
  cell{8}{1} = {r=5}{},
  vline{8,13} = {-}{0.05em},
  hline{1,13} = {-}{0.08em},
  hline{2} = {2-17}{0.03em},
  hline{3,8} = {-}{0.05em},
}
\textbf{Model} & \textbf{Sampling Ratio}     &                 &            & \textbf{80\%} &            &                 &                 &            & \textbf{90\%} &            &                 &                 &            & \textbf{100\%} &            &                 \\
               & \textbf{Ethnic Group} & \textbf{1}      & \textbf{2} & \textbf{3}    & \textbf{4} & \textbf{5}      & \textbf{1}      & \textbf{2} & \textbf{3}    & \textbf{4} & \textbf{5}      & \textbf{1}      & \textbf{2} & \textbf{3}     & \textbf{4} & \textbf{5}      \\
\textbf{LR}    & \textbf{1}            & 0.0000          & 0.1407     & 0.3530        & 0.5661     & \textbf{0.8130} & 0.0000          & 0.1424     & 0.3558        & 0.5698     & \textbf{0.8141} & 0.0000          & 0.1429     & 0.3542         & 0.5675     & \textbf{0.8120} \\
               & \textbf{2}            & 0.1407          & 0.0000     & 0.0955        & 0.3095     & 0.6872          & 0.1424          & 0.0000     & 0.0979        & 0.3141     & 0.6872          & 0.1429          & 0.0000     & 0.0966         & 0.3115     & 0.6853          \\
               & \textbf{3}            & 0.3530          & 0.0955     & 0.0000        & 0.1061     & 0.5593          & 0.3558          & 0.0979     & 0.0000        & 0.1051     & 0.5553          & 0.3542          & 0.0966     & 0.0000         & 0.1043     & 0.5541          \\
               & \textbf{4}            & 0.5661          & 0.3095     & 0.1061        & 0.0000     & 0.4005          & 0.5698          & 0.3141     & 0.1051        & 0.0000     & 0.3983          & 0.5675          & 0.3115     & 0.1043         & 0.0000     & 0.3971          \\
               & \textbf{5}            & \textbf{0.8130} & 0.6872     & 0.5593        & 0.4005     & 0.0000          & \textbf{0.8141} & 0.6872     & 0.5553        & 0.3983     & 0.0000          & \textbf{0.8120} & 0.6853     & 0.5541         & 0.3971     & 0.0000          \\
\textbf{MLP}   & \textbf{1}            & 0.0000          & 0.1421     & 0.3520        & 0.5663     & \textbf{0.8111} & 0.0000          & 0.1426     & 0.3535        & 0.5681     & \textbf{0.8118} & 0.0000          & 0.1425     & 0.3533         & 0.5679     & \textbf{0.8119} \\
               & \textbf{2}            & 0.1421          & 0.0000     & 0.0960        & 0.3120     & 0.6853          & 0.1426          & 0.0000     & 0.0964        & 0.3135     & 0.6856          & 0.1425          & 0.0000     & 0.0967         & 0.3139     & 0.6865          \\
               & \textbf{3}            & 0.3520          & 0.0960     & 0.0000        & 0.1059     & 0.5546          & 0.3535          & 0.0964     & 0.0000        & 0.1068     & 0.5548          & 0.3533          & 0.0967     & 0.0000         & 0.1066     & 0.5554          \\
               & \textbf{4}            & 0.5663          & 0.3120     & 0.1059        & 0.0000     & 0.3955          & 0.5681          & 0.3135     & 0.1068        & 0.0000     & 0.3958          & 0.5679          & 0.3139     & 0.1066         & 0.0000     & 0.3967          \\
               & \textbf{5}            & \textbf{0.8111} & 0.6853     & 0.5546        & 0.3955     & 0.0000          & \textbf{0.8118} & 0.6856     & 0.5548        & 0.3958     & 0.0000          & \textbf{0.8119} & 0.6865     & 0.5554         & 0.3967     & 0.0000          
\end{tblr}}
\end{table*}

\begin{table*}
\centering
\caption{False Positive Rates for Five Groups with Three Sampling Ratios.}
\label{tab:evenfpr}
\resizebox{0.9\textwidth}{!}{
\begin{tblr}{
  cells = {c},
  vline{5} = {-}{0.05em},
  hline{1,8} = {-}{0.08em},
  hline{2-3} = {-}{0.05em},
}
\textbf{Model}          &                          & \textbf{Logistic Regression} &                          &                          & \textbf{Multilayer Perceptron} &                          \\
\textbf{Sampling Ratio} & \textbf{80\%}            & \textbf{90\%}                & \textbf{100\%}           & \textbf{80\%}            & \textbf{90\%}                  & \textbf{100\%}           \\
\textbf{1}              & \textbf{0.1285 ± 0.0161} & \textbf{0.1367 ± 0.0135}     & \textbf{0.1383 ± 0.0144} & \textbf{0.1159 ± 0.0513} & \textbf{0.1464 ± 0.0525}       & \textbf{0.1384 ± 0.0378} \\
\textbf{2}              & 0.118 ± 0.018            & 0.1124 ± 0.0147              & 0.1127 ± 0.012           & 0.0872 ± 0.0304          & 0.1234 ± 0.0318                & 0.1067 ± 0.0306          \\
\textbf{3}              & 0.1259 ± 0.0131          & 0.1272 ± 0.014               & 0.1277 ± 0.0177          & 0.1 ± 0.026              & 0.1341 ± 0.043                 & 0.1173 ± 0.0268          \\
\textbf{4}              & 0.1232 ± 0.0175          & 0.1298 ± 0.0113              & 0.1231 ± 0.0083          & 0.0948 ± 0.0257          & 0.1327 ± 0.0342                & 0.1046 ± 0.0275          \\
\textbf{5}              & 0.0603 ± 0.0072          & 0.0624 ± 0.0058              & 0.0623 ± 0.0057          & 0.0469 ± 0.0193          & 0.0637 ± 0.0199                & 0.0533 ± 0.0175          
\end{tblr}}
\end{table*}

\section{Discrepancy for AI fairness}
As we discussed earlier, our proposed approach can be used to quantify the discrepancies in the distribution of user (sample) profiles across multiple groups. We consider that these discrepancies can negatively impact the fairness of machine learning methods. Intuitively, the training process of machine learning models may struggle to extract patterns equally from two parts of a dataset with significant discrepancies. We expect that the discrepancy can serve as a data exploratory metric to alert AI users to the risk of fairness issues and support fairness analysis in AI.

\subsection{Predefined Range Groups (Fixed Intervals)}
\label{sec:fixedintervals}
Now, we will show how discrepancies relate to potential AI bias. We selected logistic regression (LR) to classify deprivation indices for LSOAs using the Census 2021 data from previous experiments. To simplify the classification task, we redefined the deprivation indices as deprived (indices 1-5, labelled as 0) and not deprived (indices 6-10, labelled as 1). We randomly split the dataset into training and validation sets in an 8:2 ratio, and the experiments are repeated for 10 times. 

The bias is measured using the False Positive Rate (FPR). In this study, we argue that FPR deserves greater attention, as it quantifies the extent to which deprived areas are incorrectly predicted as not deprived. Such misclassification could potentially exacerbate deprivation. For instance, groups with higher FPR may not receive the deserved attention they need from the government when it comes to resource allocation decisions.

The experimental results indicate that the two groups with the largest discrepancy values exhibit the greatest difference in FPR. As shown in \Cref{tab: unevenaifairness}, the discrepancy between the 0-20\% group and the 80-100\% group is the largest with the value 0.9301. At the same time, the former group has the smallest FPR of 0, while the latter group has the largest FPR of 0.1225. This study reveals that the machine learning model can treat areas predominantly populated by non-ethnic minorities unfairly. Moreover, significant discrepancies between groups within a dataset warrant attention, as these disparities may lead to differential treatment by machine learning models.

\subsection{Equal-Size Groups (Quantile-Based)}
We observed that the grouping method employed in the aforementioned experiments presents a data imbalance issue, as the 0-20\% group constitutes 82.59\% of the LSOA samples. This imbalance may undermine the robustness of our findings, given the significant decrease in the number of samples from the 0-20\% group to the 80-100\% group. Consequently, the 80-100\% group may achieve a 0 FPR with a small number of test samples, despite having a limited amount of training data.

To address this issue, we propose an alternative grouping method. First, we calculate the ME population percentages for all LSOAs, then sort and divide them into five groups, ensuring each group contains an equal number of samples. The ME population percentage ranges for these groups are as follows: 0-0.96\%, 0.96-2.33\%, 2.33-6.16\%, 6.16-17.34\%, and 17.34-95.02\%. These groups are labeled as 1, 2, 3, 4, and 5, respectively. It is worth noting that the average ME population percentage in the UK is 18\%, indicating that only the last group can be considered representative of the ME population.

To further explore the correlations between discrepancy and AI fairness, we conducted experiments using two models: LR and a Multilayer Perceptron (MLP). Additionally, we applied two sampling ratios (90\% and 80\%) to randomly generate two new datasets. From our perspective, the sampled datasets should preserve the patterns discussed in the previous section, as random sampling does not substantially alter the overall distribution. All the results presented in \Cref{tab:evensamplingdiscrepancy} and \Cref{tab:evenfpr} are derived from experiments repeated 10 times.

In \Cref{tab:evensamplingdiscrepancy}, we observed that the discrepancies between Groups 1, 2, 3, 4, and 5 have increased compared to the values presented in \Cref{tab: unevenaifairness}. We believe this is due to all ME LSOAs being concentrated in group 5 under the new grouping method, which explains this observation. In other words, the ME group and non-ME groups have differences in their user profiles. Meanwhile, the largest discrepancy remains between group 1 (predominantly non-ME) and Group 5 (predominantly ME) across the two sampled datasets (90\% and 80\%) and the original dataset (100\%). In \Cref{tab:evensamplingdiscrepancy}, ``LR" and ``MLP" indicate that the discrepancy calculations were performed on independently sampling datasets. The datasets are used to observe the correlations between discrepancy and AI bias. 

In \Cref{tab:evenfpr}, the first group (predominantly non-ME), approximately equivalent to the 0-20\% group in \Cref{tab: unevenaifairness}, still exhibits the largest FPR across both models and all three datasets. Meanwhile, Group 5 continues to have the smallest FPR. Thus, the results align with the findings presented in \Cref{sec:fixedintervals}; the two groups with large discrepancy values may be treated differently by AI. 

Additionally, compared to the results shown in \Cref{tab: unevenaifairness}, we observe that the FPR values for Groups 2$\sim$4 increase along with the increase of discrepancy values between Groups 2$\sim$4 and Group 5. Meanwhile, the overall discrepancies within Groups 2$\sim$4 are smaller than the discrepancies between Group 5 and the other groups (Groups 1$\sim$4). This indicates that the features of the data for Groups 2$\sim$4 are relatively similar, and models treat them similarly. As shown in \Cref{tab: unevenaifairness}, Groups 1$\sim$4 received relatively similar FPRs, while the FPR for Group 5 is significantly smaller. In summary, we believe our proposed method effectively quantifies the discrepancies between different groups. Additionally, these values are important for informing AI users and highlight potential risks of AI unfairness.

\section{Conclusion and Limitations}

In conclusion, the issue of AI fairness is of paramount importance and warrants attention from all stakeholders. In our research, we addressed this challenge by focusing on quantifying the discrepancies present in data, recognising that AI models heavily rely on data for their performance. Our proposed data-driven approach is aligned with the LNOB initiative, as it aids in discovering and addressing discrepancies between user-defined groups, thus contributing to efforts to mitigate inequality. Moreover, we believe that our proposed approach holds promise for applications across a broad spectrum of tasks, offering insights to develop fair AI models. Through testing on three datasets, we have demonstrated the efficacy and informativeness of our approach, yielding satisfactory results. Our proposed approach can be considered as an approximation of bias, as selecting different parameters for LCA may yield slightly varying results, to address this we have done hyperparameter optimisation. 

In summary, our research represents a significant step towards promoting fairness in AI and offers an innovative avenue for social science research. By highlighting data-driven approaches and their alignment with broader societal initiatives, we aim to foster a more equitable and inclusive landscape for AI development and deployment.

\section*{Acknowledgement}
This work was supported by URKI’s Engineering and Physical Sciences Research Council (EPSRC) grant numbers EP/W03235X/1, EP/W032333/1, EP/W032341/1, EP/ W032058/1, EP/W032082/1 under the Protecting Minority Ethnic Communities Online (PRIME) project. The authors are also grateful to the PRIME team.

\bibliographystyle{IEEEtran}
\bibliography{bibs.bib}{}

\begin{appendices}
\renewcommand{\thefigure}{\Alph{section}.\arabic{figure}}
\renewcommand{\thetable}{\Alph{section}.\arabic{table}}
\setcounter{figure}{0}
\setcounter{table}{0}

\section{The \PRIMEproject}
\label{appendix:prime}

As part of the \textit{\PRIME}project, a multilingual online survey, available in 10 languages, was conducted to investigate the experiences of individuals from minority ethnic groups with digitalised housing, health, and energy services. The survey contains a total of 32 questions. The survey data includes 594 responses from England and 284 responses from Scotland. In terms of respondent selection, researcher carefully determined the required number of participants from each ethnic group in England using a proportional allocation method based on their respective population percentages from the 2021 Census (England \& Wales). However, due to the unavailability of Scotland's 2021 census results during the planning phase, the project aimed to limit the number of respondents from each ME group in Scotland to a maximum of 40. The total number of survey respondents was 878. A detailed breakdown of the respondents' demographic information, including their ethnicities, is provided in Table \ref{table: survey_prime}.

\begin{table}[ht]
\centering
\caption{The number of respondents from each ME group in both England and Scotland}
\label{table: survey_prime}
\resizebox{7cm}{!}{
\begin{tabular}{crrr} 
\toprule
\PRIMEprojects Target Ethnic Group       & \multicolumn{1}{c}{England} & \multicolumn{1}{c}{Scotland} & \multicolumn{1}{c}{Total} \\
\midrule
African                         & 176                         & 37                            & 213                       \\
Bangladeshi                     & 97                          & 41                            & 138                       \\
Indian                          & 93                          & 40                            & 133                       \\
Chinese                         & 63                          & 39                            & 102                       \\
Pakistani                       & 62                          & 40                            & 102                       \\
Caribbean                       & 47                          & 32                            & 79                        \\
Mixed or Multiple ethnic groups & 56                          & 55                            & 111                       \\
\midrule
Total                           & 594                         & 284                           & 878                       \\
\bottomrule
\end{tabular}}
\end{table}

\begin{figure}
    \centering
    \includegraphics[scale=0.1]{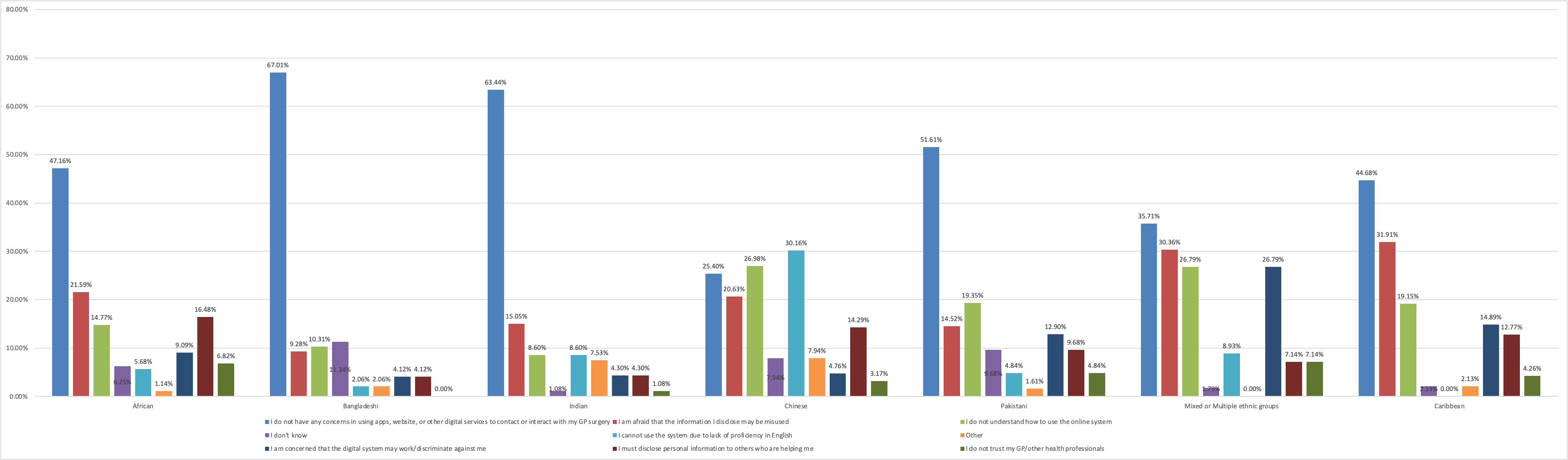}
    \caption{This figure gives an example of the exploratory data analysis (EDA) for the \textit{\PRIME}project's England data focuses on Question 21: ``Which of the following concerns do you have about communicating with your general practice (GP) through apps, websites, or other online services?" The bars represent the portion of respondents from each ethnic group selecting each option. Best viewed when zoomed in.}
    \label{fig:q20_england}
\end{figure}

We show the distribution of responses from seven ethnic groups in England regarding health and digital services based on \textit{\PRIME} project's data. We observed a notable discrepancy in the Chinese group, with 30.16\% lacking English proficiency and 26.98\% struggling to use the online system, while most other ethnic groups reported no concerns, as shown in \Cref{fig:q20_england}. 

\begin{figure}
    \centering
    \includegraphics[scale=0.2]{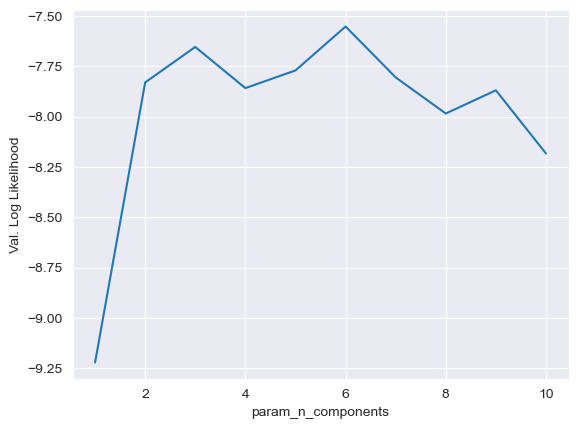}
    \caption{The example of hyperparameter optimisation process to seek the elbow point. Best viewed when zoomed in.}
    \label{fig:hpo1}
\end{figure}

It should be noted that the definitions of ethnicities differ between Scotland's census and the England \& Wales census. For instance, in the England and Wales census, individuals may select ``Pakistani" as a sub-category under the broader ``Asian or Asian British" category, whereas in Scotland census, this category is listed as ``Pakistani, Pakistani Scottish, or Pakistani British." For consistency, this paper adopts the ethnicity naming conventions used in the England and Wales census.

The questions selected to quantify discrepancies in this study include:
\begin{itemize}
    \item Question 20: Which of the following concerns do you have about communicating with your GP through apps, websites or other online services? 
    \item Question 23: Do you have any concerns about using an app, website or other digital service for these housing-related activities?
    \item Question 27: Do you have any concerns about using an app, website or digital system to carry out energy-related activities?

\end{itemize}

Regarding the selection of the number of latent classes, we used grid search for hyperparameter optimisation with 10-fold cross-validation. The search space for all experiments was adjusted based on empirical observations. In the \textit{\PRIME} project experiments, we set the range from 2 to 10, while for EVENs and Census experiments, it was set from 2 to 30. One example result of hyperparameter optimisation is shown in \Cref{fig:hpo1}.

\section{Census 2021 (England and Wales)}
\label{appendix:census}
\setcounter{figure}{0}
\setcounter{table}{0}

\begin{table}[h]
\centering
\caption{The correlation analysis between Deprivation 2019 and the results obtained from our proposed based on Census 2021.}
\label{tab:correlation}
\resizebox{0.2\textwidth}{!}{
\begin{tabular}{c|cc} 
\hline
         & Pearson & Spearman  \\ 
\hline
0-20\%   & 0.9802  & 1         \\
20-40\%  & 0.9769  & 1         \\
40-60\%  & 0.9949  & 0.9       \\
60-80\%  & 0.9829  & 1      \\
80-100\% & 0.9830  & 1         \\
\hline
\end{tabular}}
\end{table}

\begin{table}[h]
\centering
\caption{The percentages of LSOAs in different ME population groups cross 10 deprivation levels based on the Deprivation 2019 dataset and the total number LSOAs in each group.}
\label{tab:deprivcross}
\resizebox{0.5\textwidth}{!}{
\begin{tblr}{
  cells = {c},
  hlines,
  vlines,
}
         & 1       & 2       & 3       & 4       & 5       & 6       & 7       & 8       & 9       & 10      & Total LSOAs \\
0-20\%   & 8.62\%  & 7.83\%  & 8.28\%  & 9.14\%  & 10.04\% & 10.49\% & 11.03\% & 11.32\% & 11.52\% & 11.75\% & 26273       \\
20-40\%  & 12.22\% & 18.93\% & 18.12\% & 14.81\% & 10.77\% & 8.05\%  & 5.96\%  & 4.04\%  & 3.79\%  & 3.31\%  & 3592        \\
40-60\%  & 20.10\% & 23.19\% & 17.49\% & 13.02\% & 8.71\%  & 6.18\%  & 4.48\%  & 3.58\%  & 2.12\%  & 1.14\%  & 1229        \\
60-80\%  & 31.13\% & 22.67\% & 16.58\% & 12.35\% & 8.12\%  & 4.91\%  & 2.71\%  & 1.18\%  & 0.34\%  & 0.00\%  & 591         \\
80-100\% & 41.60\% & 26.40\% & 20.00\% & 7.20\%  & 1.60\%  & 2.40\%  & 0.80\%  & 0.00\%  & 0.00\%  & 0.00\%  & 125         
\end{tblr}}
\end{table}

\Cref{tab:correlation} provides details of the correlation analysis discussed in \Cref{sec: census2021correlation}. Meanwhile, \Cref{tab:deprivcross} presents the distribution of the deprivation index across 10 predefined LSOA groups.

\section{Experiment Details}
\label{appendix:experimentdetails}

It is worth noting that our approach is not time-consuming like deep learning models. The time required, based on the hardware shown in \Cref{tab:hardware}, ranges from 5 seconds to a maximum of 5 minutes, depending on the dataset volume.

\setcounter{figure}{0}
\setcounter{table}{0}
\begin{table}[h]
\centering
\caption{The hardware and software details of experiments.}
\label{tab:hardware}
\begin{tabular}{|c|c|}
\hline
\textbf{}       & \textbf{Hardware}                                \\ \hline
\textbf{CPU}    & 12th Gen Intel(R) Core(TM) i9-12950HX   2.30 GHz \\ \hline
\textbf{GPU}    & NVIDIA GeForce RTX 3080 Ti Laptop GPU            \\ \hline
\textbf{Memory} & 1TB                                              \\ \hline
\textbf{RAM}    & 64.0 GB                                          \\ \hline
\textbf{OS}     & Windows 11 Pro                                   \\ \hline
\end{tabular}
\end{table}
\end{appendices}

\end{document}